\def\year{2017}\relax
\pdfoutput=1
\documentclass[letterpaper]{article}
\usepackage{aaai17}
\usepackage{times}
\usepackage{helvet}
\usepackage{courier}
\usepackage{color}
\usepackage{dsfont}
\usepackage{amsmath}
\usepackage{amssymb}
\usepackage{graphicx}
\usepackage{array, booktabs}
\usepackage{subfigure}
\usepackage{url}
\usepackage{threeparttable}
\usepackage[]{algorithm2e}
\newcommand{\citenoun}[1]{{\citeauthor{#1} \shortcite{#1}}}
\newcommand{\cut}[1]{{}}
\newcommand{\outline}[1]{{}}
\newcommand{\x}{\mathbf{x}}

\begin{document}
%
\nocopyright
\title{ Identifying leading indicators of product recalls from online reviews using positive unlabeled learning and domain adaptation}
\author{
  Shreesh Bhat \and Aron Culotta\\
  Department of Computer Science\\
  Illinois Institute of Technology\\
  Chicago, IL 60616\\
  skumarab@hawk.iit.edu, aculotta@iit.edu\\
}
\maketitle

\begin{abstract}
    \begin{quote}
Consumer protection agencies are charged with safeguarding the public from hazardous products, but the thousands of products under their jurisdiction make it challenging to identify and respond to consumer complaints quickly. From the consumer's perspective, online reviews can provide evidence of product defects, but manually sifting through hundreds of reviews is not always feasible. In this paper, we propose a system to mine Amazon.com reviews to identify products that may pose safety or health hazards. Since labeled data for this task are scarce, our approach combines positive unlabeled learning with domain adaptation to train a classifier from consumer complaints submitted to the U.S. Consumer Product Safety Commission. On a validation set of manually annotated Amazon product reviews, we find that our approach results in an absolute F1 score  improvement of 8\% over the best competing baseline. Furthermore, we apply the classifier to Amazon reviews of known recalled products; the classifier identifies reviews reporting safety hazards prior to the recall date for 45\% of the products. This suggests that the system may be able to provide an early warning system to alert consumers to hazardous products before an official recall is announced.
    \end{quote}
\end{abstract}

\section{Introduction}

The U.S. Consumer Product Safety Commission (CPSC), created by the
1972 Consumer Product Safety Act, is ``charged with protecting the
public from unreasonable risks of injury or death associated with the
use of the thousands of types of consumer products under the agency's
jurisdiction. Deaths, injuries, and property damage from consumer
product incidents cost the nation more than \$1 trillion
annually.''\footnote{https://www.cpsc.gov/About-CPSC} Typically, the
CPSC learns about hazardous products through consumer reports, either
through a phone hot-line, or through their online portal
SaferProducts.gov.

For a subset of these complaints, the CPSC may decide that action is
warranted, which most commonly takes the form of a ``cooperative
recall,'' in which the manufacturer agrees to issue a voluntary recall
based on the CPSC's findings. As the agency notes, ``due to the large
volume of reports received by the CPSC each year, agency staff,
unfortunately, cannot investigate and respond to every report on an
individual
basis.''\footnote{https://www.cpsc.gov/About-CPSC/Contact-Information}
For example, in FY2015, the CPSC completed 410 cooperative recalls; by
comparison, the CPSC received 85,000 calls to their hot-line and 2,539
incidents submitted to their online
portal.\footnote{https://www.cpsc.gov/s3fs-public/FY15AnnualReport.pdf}

Given the large number of products under its jurisdiction, the CPSC faces a number of regulatory challenges:
\begin{itemize}
\item Triage: Given the many potential hazards to investigate, how should they be prioritized?
\item Discovery: How can new product hazards be efficiently reported to the CPSC?
\item Notification: The time lag from report to recall can span multiple months (involving investigation and negotiations with the firm). How can consumers be notified more quickly of a potential hazard?
\end{itemize}

In this paper, we propose a system to help with these tasks by
identifying product reviews on Amazon.com that indicate a potential
safety or health hazard. The resulting system helps with discovery by
identifying hazards that may not be submitted to the CPSC directly; it
helps with triage by enabling complaints to be aggregated to identify
high priority products; and it helps with notification by enabling
consumers to be alerted immediately when hazardous reviews are posted
on Amazon.

To train the text classification system, we use consumer complaints
data uploaded to the CPSC portal at {\tt ConsumerSafety.gov}. These
``positive'' instances are combined with thousands of unlabeled
instances from Amazon.com reviews using Positive Unlabeled
Learning~\cite{li2005learning}. However, standard training algorithms
underperform on this task, because these two data sources differ in
systematic ways. To deal with this issue, we build on work in learning
under dataset shift~\cite{heckman1977sample,zadrozny2004learning} to
train a more accurate classifier. The resulting classifier identifies
reviews mentioning safety hazards with an F1 score of 84\%, an
absolute improvement of 8\% over the best baseline.  Furthermore, we
applied the classifier to reviews of known recalled products, and
found that for 45\% of the products, the system detected a review
reporting a health or safety hazard prior to the recall date. This
suggests that the system may be able to provide an early warning
system to alert consumers to potentially hazardous products.






\section{Data}

Our goal is to build a text classifier to determine whether a product
review on Amazon.com reports a potential safety or health hazard of a
product. As we expect such reviews to be rare, it is difficult to
construct a training set in the traditional way of annotating a random
sample of reviews. Instead, we consider the consumer complaints
database on the CPSC website {\tt SaferProducts.gov}. We supplement
this with a large set of unlabeled Amazon reviews to build the
classifier using Positive Unlabeled learning. For validation, we
consider two additional data sources: a small set of annotated Amazon
reviews, and a set of products that were recalled by the CPSC over the
past 10 years. Below, we describe these data in more detail.

\subsection{CPSC Complaints Database}
The Consumer Product Safety Improvement Act was passed in 2008 to
strengthen the CPSC by increasing its budget and expanding its
regulatory tools. In addition, the law mandated that the CPSC create a
publicly searchable database of consumer submitted reports of
hazardous products. The website was launched in 2011 as {\it
SaferProducts.gov}. To reduce the number of false reports, the site
requires information about the consumer, photographs to document the
report, and information about the specific product and
manufacturer. Once a report is submitted, it is vetted by the CPSC,
and, if deemed valid, it is first sent to the manufacturer for
comment. Thus, it can take approximately 15 business days for a report
to appear on the website (though longer times are possible, depending
on volume and capacity).

For this paper, we focus on children products, since these tend to be
the most vulnerable to health and safety hazards. We collected 2,010
complaints from the ``Babies \& Kids'' category from {\tt
SaferProducts.gov}, from March 2011 -- May
2016. Table~\ref{tab.complaint.types} shows the top five most frequent
product types in this data.

Each report has an incident description, which ranges from 4 to 1,683
words (median=98). Two short example descriptions are below:
\begin{quote}
I went to change the sheets on my son's crib, and the mattress had
broken in the middle. Plastic was all over the mattress. Springs are
close to poking thru.
\end{quote}
\begin{quote}
A piece of the toy fall out and my daughter almost swallow it. we have
to put out hand in her mouth to take it back. My daughter cannot
breathe for a second.
\end{quote}

\begin{table}[t]
\centering
\begin{tabular}{|c|c|}
  \hline
  \textbf{Product Type} & \textbf{Count}\\
  \hline
  Cribs & 407\\
  Bassinets or Cradles & 258\\
  Diapers & 209\\
  Pacifiers or Teething Rings & 186\\
  Baby Exercises & 132\\
\hline
\end{tabular}
\caption{Top 5 product types from the ``Babies \& Kids'' category in the consumer complaint database {\tt SaferProducts.gov} (of 2,010 total complaints).\label{tab.complaint.types}}
\end{table}

When training the classifier, we assume that these 2,010 incident
descriptions are positive examples (i.e., indicative of health or
safety hazard). We refer to this as the {\bf complaints database}.

\subsection{Amazon Product Reviews}
We collect 915,446 Amazon reviews in the ``Baby'' category from the
dataset introduced in \citenoun{mcauley2015image}, from August 2008 -
July 2014.\footnote{{\tt http://jmcauley.ucsd.edu/data/amazon/}} These
reviews range from 1 to 4,546 words (median=55). We refer to this as
the {\bf reviews database}.

\subsection{Labeled Review Data}

For validation, we manually annotated 448 Amazon reviews as to whether
they report a hazardous or unsafe product. To construct this data, we
combined uniform sampling with keyword search to identify possible
positive examples (e.g., terms like ``hurt'' and ``dangerous''). The
final dataset contains 97 positive (hazardous) reviews and 351
negative (non-hazardous) reviews.

Table~\ref{tab.labeled.data} shows five examples (three positive, two
negative). A key challenge is distinguishing between reviews
indicating a safety hazard and reviews that indicate more benign
faults of the product. We refer to this as the {\bf validation data}.

\begin{table}[t]
\centering
\begin{tabular}{|c|p{6.5cm}|}
  \hline
  \textbf{Label} & \textbf{Review snippet}\\
  \hline
  1 & ``This item needs to be taken off the market.  My son almost suffocated to death in this...''\\
\hline
  1 & ``I had this product 2 hours and the leg snapped. My baby rolled forward and hit his head. It is now in the trash!!''\\
\hline
   1 & ``...When I was cleaning the tray, my daughter leaned forward and the whole chair with booster seat fell down. My daughter got a bump on her head...''\\
\hline
\hline
  0 & ``I'm sending this product back today!!! I thought the Recaro booster seat would be light weight and trendy.  This seat was so heavy I could hardly get it out of the box.''\\
\hline
 0 & ``It's cheaply made. I washed it on the gentle cycle and it began to fall apart :(``\\
\hline
\end{tabular}
\caption{Example reviews labeled as consumer safety concern (1) or not (0).\label{tab.labeled.data}}
\end{table}

\subsection{Recall Database}
\label{sec.recall.db}
Finally, to explore the practical impact of this classifier, we
collected a set of products that were recalled by the CPSC and had
reviews in the reviews database. To do so, we first collected 6,741
recalled products from {\tt cpsc.gov}\footnote{API:
https://www.cpsc.gov/Recalls/CPSC-Recalls-Application-Program-Interface-API-Information/}. We
used a semi-automated process to match each recalled product with an
Amazon product in the reviews database. To do so, we first filtered
the recalled product list to those containing keywords relevant to the
``Baby'' category in Amazon: stroller, car seat, crib, child carrier,
bath seat, infant carrier, bassinet, pacifier, rattle, swing, walker,
dresser. This matched 482 of the original 6,741 recall records. We
then extracted the product and/or company name from the {\tt Title}
field of the recall record, then returned Amazon products that matched
at least two terms from the product and/or company name. This
identified 3,523 products that partially matched one of 290 recall
records. Finally, we manually verified the matches, resulting in a set
of 137 Amazon products that matched one of 47 recall records (some
recalls affect multiple products). Note that none of these recalled
products were available on Amazon at the time of this writing; the
historical reviews database allows us to identify reviews written
before the product was taken off the market.

As this filtering makes clear, recalls are relatively rare events, so
the data sparsity poses a challenge for typical machine learning
training and validation workflows. This motivates our use of the
complaints database to identify reviews indicating hazardous products.

\begin{figure}[t]
  \centering
  \includegraphics[width=0.5\textwidth]{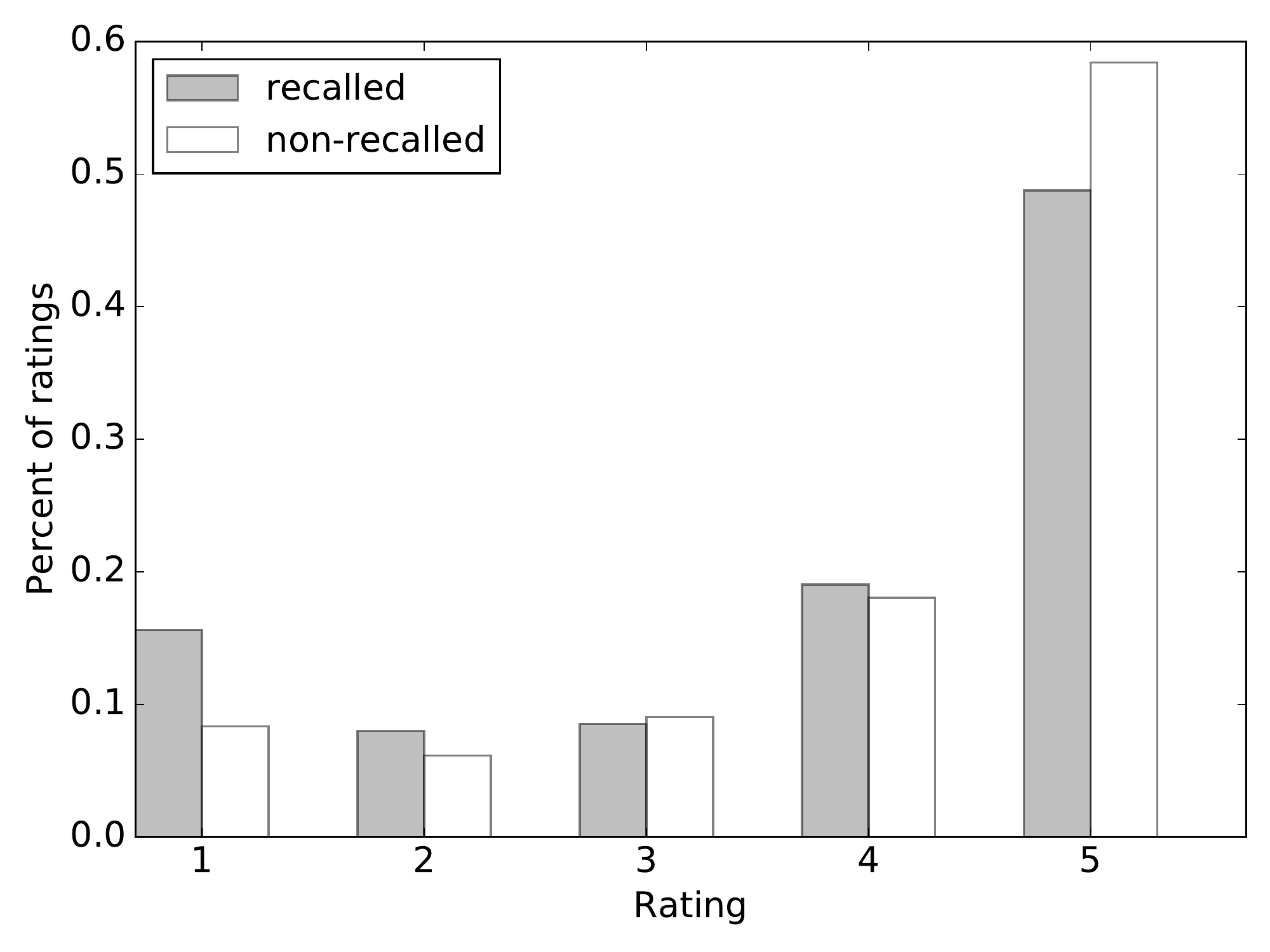}
  \caption{\label{fig.ratings} Amazon star rating distribution for reviews of recalled versus non-recalled products.}
\end{figure}

Furthermore, it is worth noting that even recalled products can have
many positive reviews. Figure~\ref{fig.ratings} shows the distribution
of star ratings for the recalled products compared with the
non-recalled products. While recalled products have a slightly lower
average rating than non-recalled products (3.8 vs. 4.1), nearly half of
the reviews for recalled products have five stars. This suggests that
using ratings alone is insufficient to identify hazardous products. We
believe this is in part due to the fact that product defects may only
affect a small subset of consumers, either due to the manufacturing
process or the way in which the consumer uses the product.
(It is also possible that fake reviews are having an impact here~\cite{mukherjee2012spotting}).

\section{Methods}

Our goal is to train a text classifier using the consumer complaints
data to classify Amazon reviews as indicative of a hazardous product
or not. We have as input a set of positive examples from the
complaints database and a set of unlabeled examples from the reviews
database. Let $\x_i\in \mathbb{R}^k$ be the $k$-dimensional feature
vector representing document $i$ and $y_i\in \{0,1\}$ be its class
label, where 1 indicates a hazardous review. Then our input consists
of a positively labeled dataset $L=\{(\x_1, 1) \ldots (\x_n, 1)\}$ of
consumer complaints and an unlabeled dataset $U=\{x_1 \ldots x_m\}$ of
Amazon reviews.

This setting can be viewed as an instance of Positive Unlabeled
learning (PU Learning ~\cite{li2005learning}), since the training set
consists of only positive and unlabeled instances. Below, we describe
a simple baseline approach to this problem, identify a problem with
this approach, then propose a new method that addresses this problem.

\subsection{Baseline method}
\label{sec.baseline}

A simple approach to PU Learning is to assume that the unlabeled
dataset $U$ contains only negative examples; i.e.,
$U \triangleq \{(\x_1, 0) \ldots (\x_m, 0)\}$. Of course, the
unlabeled data may indeed contain positive examples; however, in our
setting, hazardous reviews are rare in the reviews data, and so we
expect the amount of label noise introduced to be low.

Furthermore, in this review domain, we also have the star rating of
each review, which we can use to reduce the incidence of positive
examples incorrectly annotated as negative examples in our training
set. We expect reviews indicating safety hazards to have a low star
rating. (While Figure~\ref{fig.ratings} shows that recalled products
can have high {\it average} ratings, we expect individual reviews
mentioning health hazards to have low star ratings.)  So, we introduce
a threshold $\tau$ when sampling negative examples from the unlabeled
data; only instances with star rating greater than or equal to $\tau$
are selected. We also use a second parameter $s$ indicating the number
of negative examples to sample during training.

We use logistic regression with L2 regularization as the baseline
classifier. To handle class imbalance (there are many more negative
examples than positive examples), we weight each instance inversely
proportional to its class frequency. Thus, if there are $p$ positive
examples and $n$ negative examples, each positive example receives
weight $\frac{n + p}{2p}$, and each negative example receives weight
$\frac{n + p}{2n}$. We use Scikit-learn's LogisticRegression
implementation~\cite{pedregosa2011scikit}.

\subsection{Proposed method: Informed prior}

In addition to the small amount of label noise introduced by the
baseline method (positive examples labeled as negative), there is
another, potentially more serious difficulty with the approach for
this data. The problem stems from the selection bias in how the
positive and negative examples are collected. Specifically, certain
types of products like cribs, diapers, and night lights are
over-represented in the complaints data relative to their prevalence
in the reviews data. This leads to the inflation of coefficients
related to these products --- indeed, in the experiments below, we find
that the terms ``crib,'' ``pampers,'' and ``night light'' are among
the top ten highest weighted coefficients for the positive class for
the baseline classifier. This can lead to a number of false positives,
in which reviews of these types of products are erroneously labeled as
hazardous.

If fine-grained product subtype information were available, we could
apply standard adjustments to account for this sampling bias (e.g.,
survey weights~\cite{gelman2007struggles} or propensity
scores~\cite{rosenbaum1983central}). However, we do not have this
product subcategory information for each complaint, nor would the
schemas be equivalent between the complaints and review data.

Instead, we build on work in learning under dataset
shift~\cite{heckman1977sample,zadrozny2004learning} and
semi-supervised domain
adaptation~\cite{blitzer2006domain,kumar2010co,chen2011co}. Our
approach modifies the feature representation so that terms that are
strongly predictive of the positive class in the unlabeled dataset
have larger feature values than terms that are less predictive. Of
course, we do not know the true labels in the unlabeled data; we
instead use the baseline classifier to estimate them.

Our approach begins by fitting the baseline classifier, as defined in
the previous section. Recall that the baseline classifier with
parameters $\tau$ and $s$ constructs a training set containing all
complaint data as positive examples and a random sample of $s$ Amazon
reviews with a rating of at least $\tau$ as negative examples; we
refer to this as the {\bf baseline training set}. We then apply the
classifier trained on the baseline training set to predict the labels
for all unlabeled reviews in the Amazon review data; we refer to this
as the {\bf predicted reviews data}. Based on the examples above
(``crib'', ``pampers,'' etc.), the key observation of our approach is
that certain word features may be strongly associated with the
positive class in the original training data, but may be weakly
associated with the positive class in this predicted reviews data. For
example, in one experiment below with $\tau=5$ and $s=20,000$, we find
that in the baseline training set, 91\% of documents with the term
``pampers'' were annotated as positive examples (i.e., were from the
complaints data). However, in the predicted reviews data, only 2\% of
documents containing the term ``pampers'' were predicted to be
positive examples by the baseline classifier. So, our motivation is to
use the feature statistics in the predicted reviews data to better
inform the classifier trained on the baseline training
set. Specifically, we want to increase the importance of features that
are strongly associated with the positive class in the predicted
reviews data. We do this by modifying the value for features
proportional to their class conditional probability in the predicted
reviews data, as described next.

In order to formalize this intuition, we must introduce some
notation. Let $\hat{U} = \{(\x_1, \hat{y}_1) \ldots
(\x_m, \hat{y}_m)\}$ be the predicted reviews data; i.e., all Amazon
reviews and the corresponding class labels predicted by the baseline
classifier. Let $\theta_j \in \mathbb{R}$ be the coefficient in the
baseline model associated with word feature $j$, and let
$x_i^j \in \{0, 1\}$ be the binary feature value for feature $j$ in
document $i$. For each term feature, we compute the smoothed class
conditional probability according to the predictions in $\hat{U}$. Let
$n_{jc}$ be the number of documents containing feature $j$ that have
been assigned label $c$ by the baseline classifier:
$$
n_{jc} = \sum_{(\x_i, \hat{y}_i) \in \hat{U}} \mathds{1}[\hat{y}_i == c \wedge x_i^j==1]
$$
Then we can define the conditional probability with Laplacian smoothing as:
$$
p(y=1|x^j=1) = \frac{1 + n_{j1}}{2 + n_{j1} + n_{j0}} \triangleq p_{j1} 
$$
and similarly for $p_{j0}$ for class 0.

Let $F^+$ be the set of features with positive coefficients in the
baseline classifier, and $F^-$ be the set of features with negative
coefficients in the baseline classifier. We will use $p_{j1}$ to
transform the feature values for $F^+$, and $p_{j0}$ to transform the
feature values for $F^-$. In order to have the transformation be in
the same scale for each class, we first normalize the conditional probabilities to
sum to one for each class:
$$
\hat{p}_{j1} = \frac{p_{j1}}{\sum_{j' \in F^+}p_{j'1}}
$$
and similarly for $\hat{p}_{j0}$. To construct suitable feature
values, we want to shift these values to have a mean of 1 and be
non-negative, which we can do by multiplying each value by a constant
factor $\rho$, the ratio of the number of features to the sum of the
values $\hat{p}_{jc}$:
$$\rho = \frac{|F^+| + |F^-|}{\sum_{j \in F^+}\hat{p_{j1}}
+ \sum_{j \in F^-}\hat{p_{j0}}}$$
Finally, for all instances in the training and unlabeled data, we
replace the value of feature $j$ with the factor $(\rho
* \hat{p}_{j1})$ if $j \in F^+$ or with $(\rho * \hat{p}_{j0})$ if
$j \in F^-$.

As an example from one of the experiments below, the feature value for
the bigram ``very dangerous'' is increased to 17.4, because 29\% of
documents in the unlabeled data containing ``very dangerous'' were
classified as positive by the baseline classifier, the second highest
rate of all features. Conversely, the term ``crib'' only has a feature
value of 2.1, because only 3\% of documents in the unlabeled data
containing ``crib'' were classified as positive. This is particularly
notable given that the baseline model assigns a higher coefficient to
``crib'' (1.34) than to ``very dangerous'' (0.55).

\begin{table*}[t]
\centering
\begin{tabular}{llllll}
\toprule
          Model & Review Threshold ($\tau$) &           ROC AUC &               F1 &        Precision &           Recall \\
\midrule
 informed prior &              5 &  {\bf 97.0} $\pm $ 0.10 &  {\bf 84.3} $\pm$ 0.42 &  85.8 $\pm$ 0.90 &  {\bf 82.8} $\pm$ 0.28 \\
 informed prior &              4 &  96.4 $\pm $ 0.19 &  82.7 $\pm$ 0.47 &  86.8 $\pm$ 0.35 &  79.0 $\pm$ 0.56 \\
 informed prior &              3 &  96.3 $\pm $ 0.09 &  82.1 $\pm$ 0.41 &  {\bf 87.6} $\pm$ 0.16 &  77.3 $\pm$ 0.84 \\
       baseline &              5 &  96.1 $\pm $ 0.08 &  75.3 $\pm$ 0.43 &  72.8 $\pm$ 0.57 &  78.0 $\pm$ 0.28 \\
       baseline &              4 &  95.9 $\pm $ 0.01 &  74.8 $\pm$ 0.36 &  73.7 $\pm$ 0.76 &  75.9 $\pm$ 0.28 \\
       baseline &              3 &  95.7 $\pm $ 0.06 &  76.4 $\pm$ 0.54 &  78.4 $\pm$ 0.86 &  74.6 $\pm$ 0.28 \\
       baseline &             none &  94.0 $\pm $ 0.05 &  70.0 $\pm$ 0.24 &  79.0 $\pm$ 0.95 &  62.9 $\pm$ 0.97 \\
\bottomrule
\end{tabular}
\caption{Comparison of the baseline classifier with our informed prior method on the validation data (with standard errors).\label{tab.results}}
\end{table*}

While it may seem that changing the feature value will not affect the
final classifier, recall that we are using L2 regularization, and that
we do not standardize the feature values prior to training. Thus, all
else being equal, the regularization penalty will be effectively
smaller for features with large values than for features with small
values. Furthermore, because we optimize logistic regression with
gradient descent, the gradients for features with large values will
tend to be larger than for features with small values, causing
features with high values to be updated faster than others.

\section{Experimental Results}
\label{sec:experiments}

\begin{figure}[t]
  \centering \includegraphics[width=0.45\textwidth]{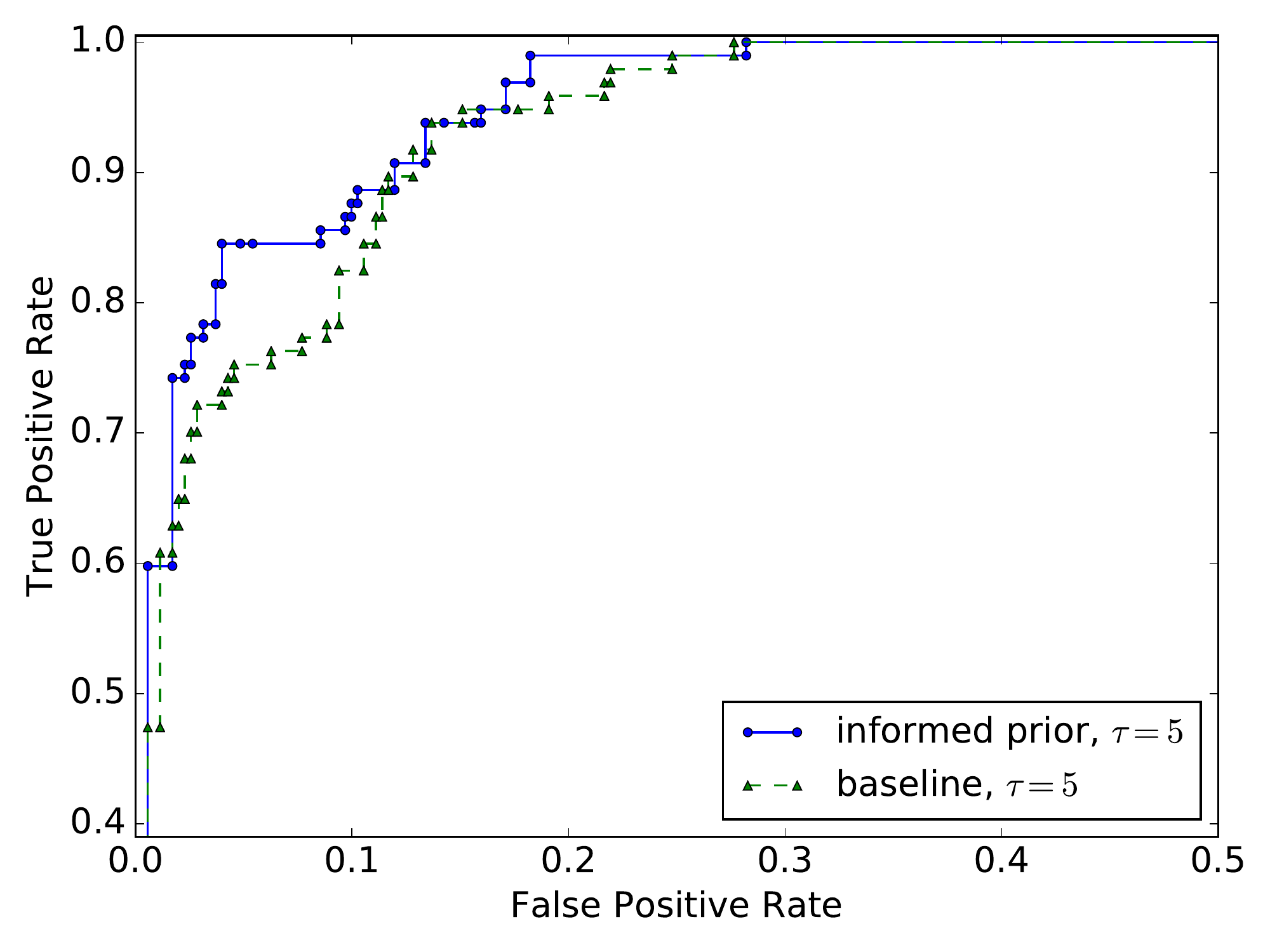} \caption{ROC
  curves for top informed prior and baseline classifiers from
  Table~\ref{tab.results}.\label{fig.roc}}
\end{figure}

\begin{table}[t]
\centering
\begin{tabular}{|p{1.5cm}|p{6.4cm}|}
  \hline
  \textbf{Model} & \textbf{Top terms}\\
  \hline
  \textbf{Informed prior} & very dangerous, cpsc, mold, smacked, swallow it, emergency room, recalled, recall, was playing, hazard, is unsafe, snapped, leaned forward, the consumer, got stuck, was hanging, burnt, injured, exploded, was chewing\\
\hline
  \textbf{Baseline} & mold, {\it pampers}, fell, {\it crib}, rock, dangerous, {\it night light}, hazard, broke, happened, {\it gate}, rash, {\it light}, recall, {\it model}, stuck, unsafe, caused, noticed, choking\\
\hline
\end{tabular}
\caption{Top 20 terms for two models. \label{tab.terms}}
\end{table}

In the experiments below, we investigate four questions:
\begin{enumerate}
\item How does our informed prior approach compare to the baseline classifier?
\item How do the parameters $\tau$ (the star rating threshold) and $s$ (the number of unlabeled examples used for training) affect accuracy?
\item How often does the classifier identify potential product hazards {\bf before} a recall is issued for a product?
\item How well does the classifier prioritize products to be investigated for hazards?
\end{enumerate}

\begin{figure}[t]
  \centering \includegraphics[width=0.45\textwidth]{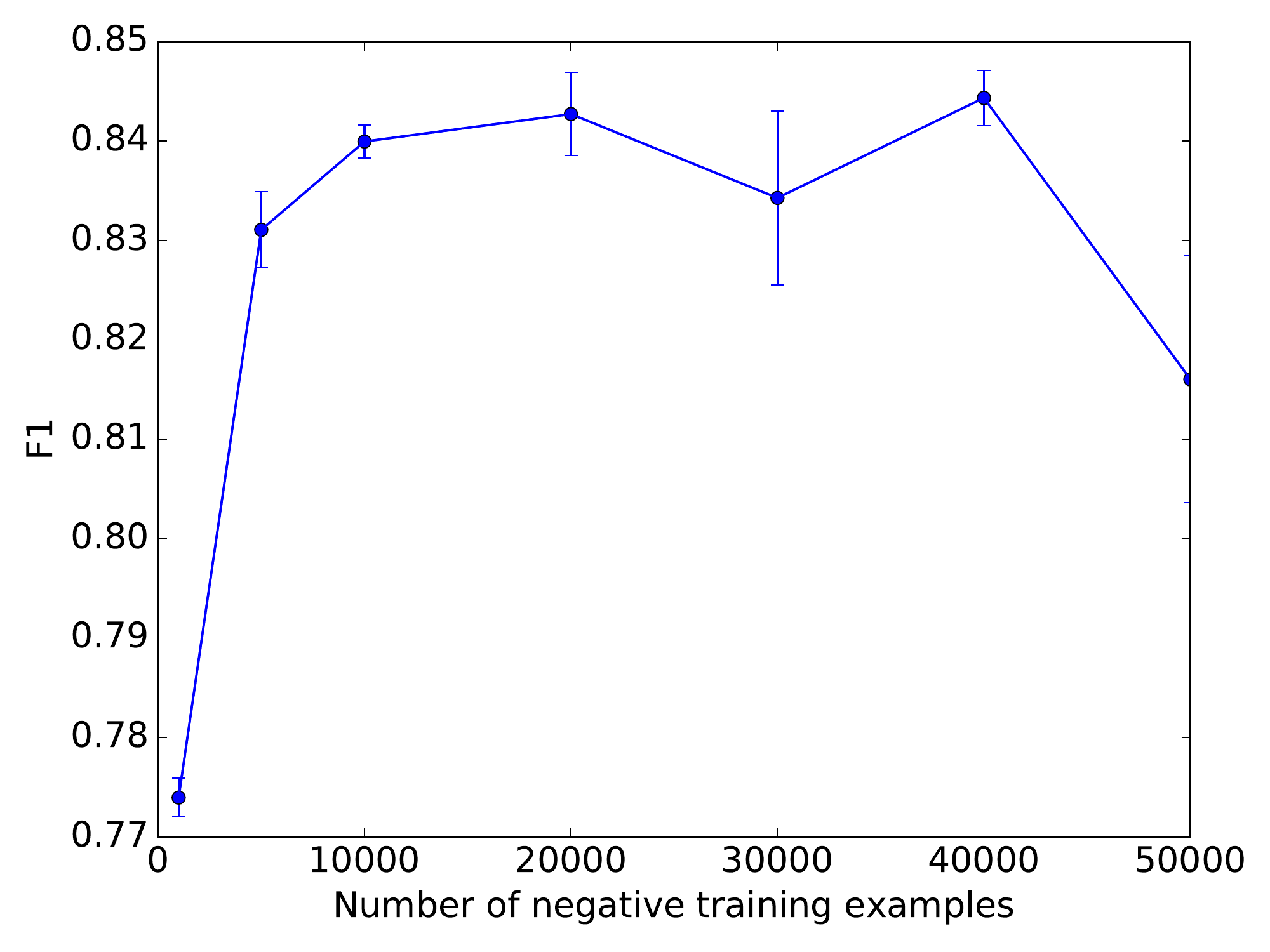} \caption{F1
  of the informed prior classifier ($\tau=5)$ as the number of
  negative examples sampled for training ($s$) increases (standard
  error bars computed from three trials).\label{fig.nneg}}
\end{figure}

\begin{figure}[t]
  \centering
  \includegraphics[width=0.5\textwidth]{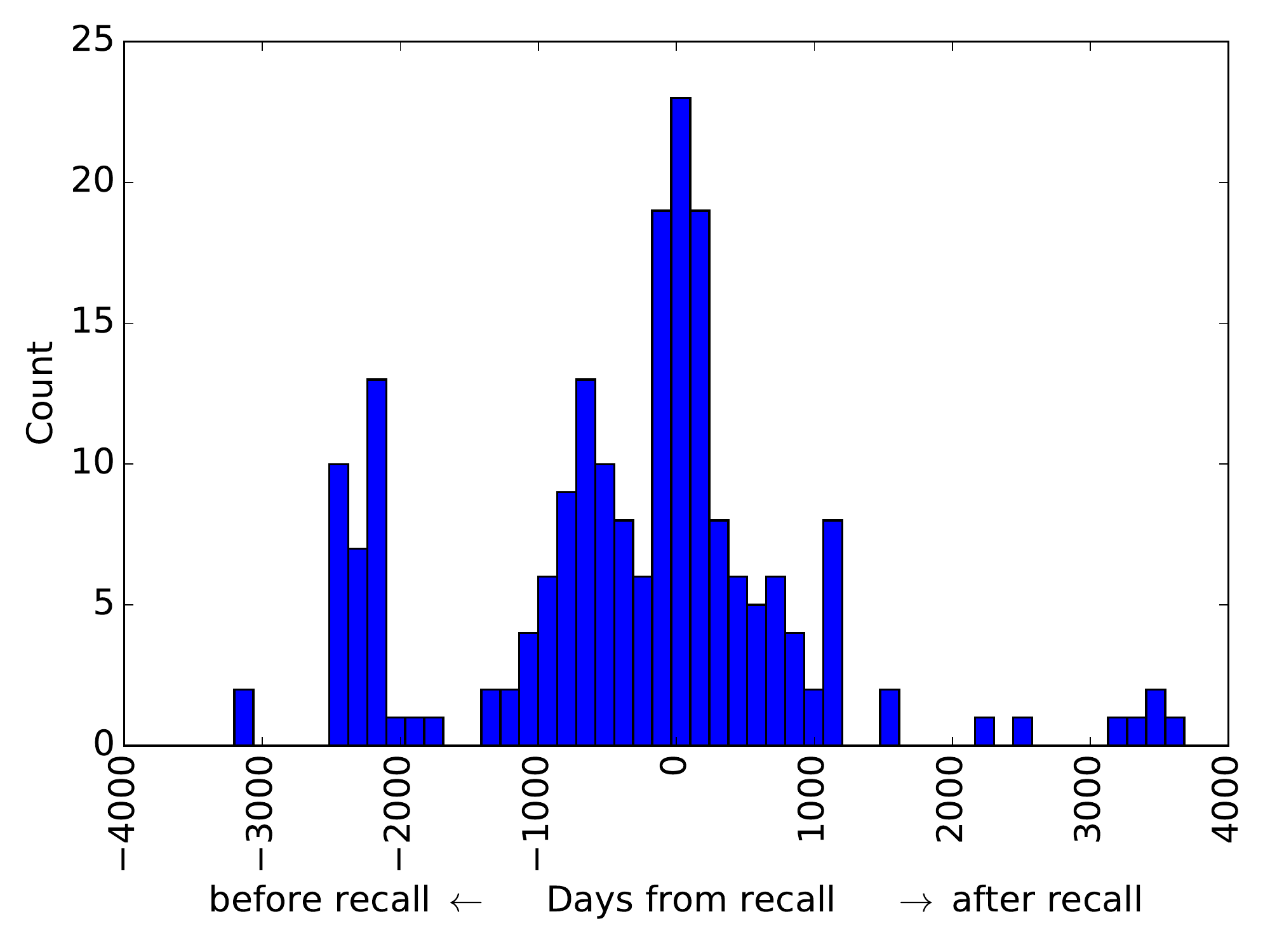}
  \caption{Histogram of when each of the identified hazardous reviews was submitted to Amazon relative to the date that the product was recalled. \label{fig.dayshist}}
\end{figure}

\begin{figure}[t]
  \centering
  \includegraphics[width=0.5\textwidth]{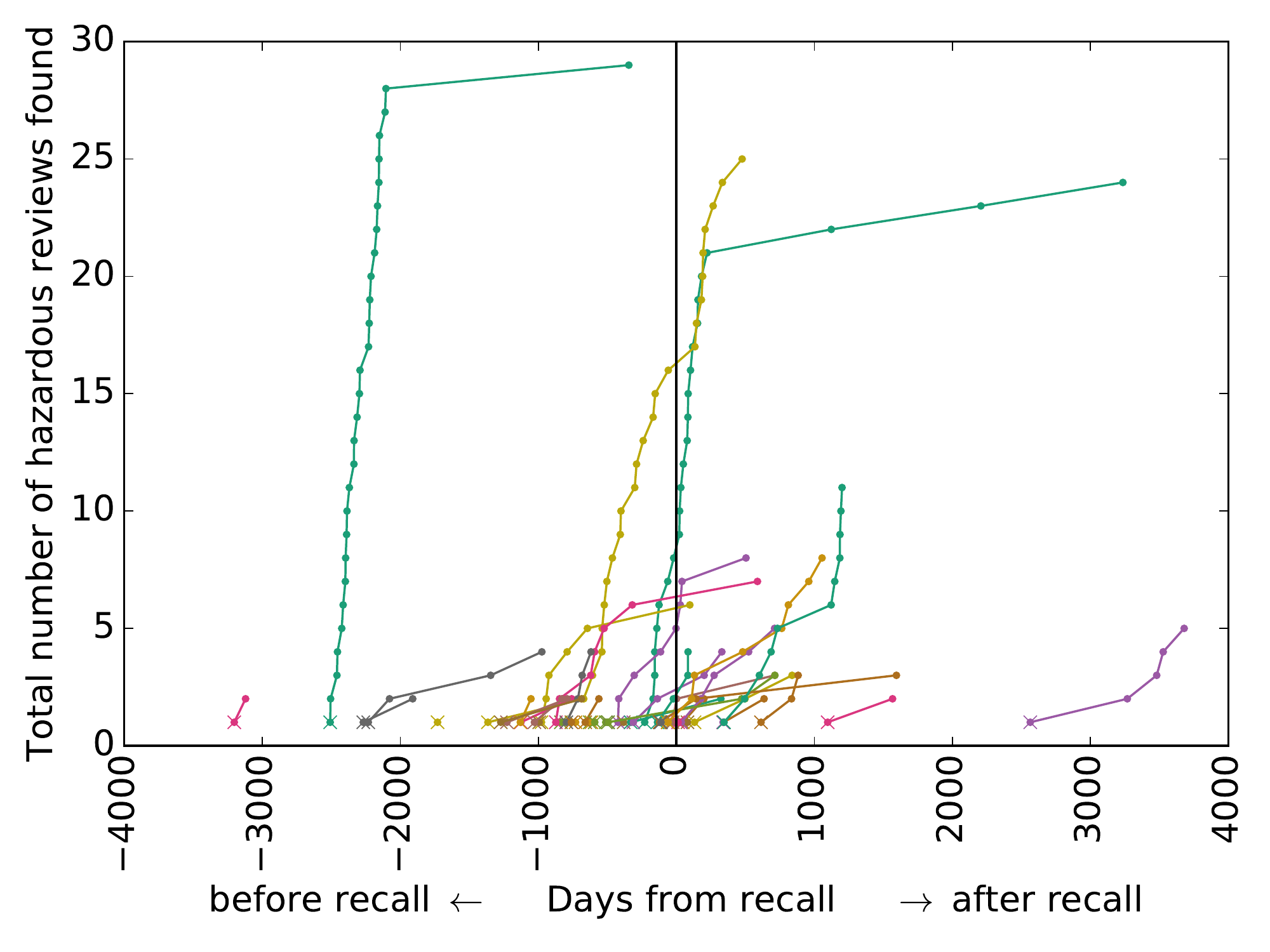}
  \caption{The number of hazardous reviews found for each recalled product over time (one line per product).\label{fig.leadtime}}
\end{figure}

We first tokenize all 915,446 Amazon reviews, retaining unigrams and
bigrams that appear in at least 50 reviews and no more than 95\% of
all reviews, resulting in 136,160 total features. We represent each
review as a binary feature vector. Using this pruned feature set, we
then vectorize the 2,010 messages in the complaints dataset, as well
as 448 labeled Amazon reviews in the validation data. For evaluation
we use Precision/Recall/F1 as well as the area under the ROC curve.

Since the baseline training set is constructed by sampling $s$ random
reviews with rating $\ge \tau$ from the unlabeled Amazon review data,
we average the results of three trials when reporting results below.

{\bf How does our informed prior approach compare to the baseline
classifier?} The primary classification results on the validation data
are shown in Table~\ref{tab.results}. Here, we fix $s=20,000$ (we will
explore it more below). We observe that across all performance
measures the informed prior method produces more accurate results than
the baseline. These results are also reinforced by the ROC curves in
Figure~\ref{fig.roc}, displaying the results for the top informed
prior and top baseline classifiers.

To better understand these results, we report in Table~\ref{tab.terms}
the 20 terms with the highest positive coefficients for the informed
prior and baseline classifiers, using $s=20,000$ and $\tau=5$. We can
see that the baseline model has many words that are likely due to
sampling bias (italicized in the table), such as ``pampers'',
``crib,'' ``night light,'' ``gate,'' and ``model.'' On the other hand,
the informed prior model gives higher weight to features such as
``very dangerous,'' ``emergency room,'' and ``is unsafe.'' Recall that
both models are fit using the same training instances; the only
difference is that feature values are increased for terms estimated to
be predictive of the positive class in the unlabeled data. We also
note that terms like ``cpsc'' and ``recalled'' arise from that fact
that some reviews either discuss a pending or past recall of a
product, or indicate that they have concurrently posted a complaint to
the CPSC database.

When we manually examine the remaining errors of the classifier, we
notice a few themes. For false positives, we observe that some reviews
describe hazards of {\it other} products, as a way to emphasize the
quality of the product being reviewed. A deeper syntactic analysis of
the reviews may be able to identify such cases. For false negatives,
we observe some reviews in which the consumer thinks there is a safety
hazard, but did not experience a first-hand injury (e.g., ``it seems
flimsy'' or ``the setup is rickety''). Our classifier is perhaps too
conservative in these instances, instead relying on more serious
reports of injury.

{\bf How do the parameters $\tau$ and $s$ affect accuracy?}
Table~\ref{tab.results} also lists results as we change $\tau$, the
review threshold used when sampling examples from the unlabeled data
to serve as negative training instances. We can see that increasing
this threshold can greatly improve the recall of the classifier, while
sometimes reducing precision. However, the overall AUC increases as
$\tau$ increases. We conjecture that the boost in recall is in part
because by removing reviews with low ratings, we remove from the
training set reviews with negative sentiment that are labeled as
non-hazardous reviews. Examples include the final two rows in
Table~\ref{tab.labeled.data}, which are critical of the product, but
do not report a specific safety concern. Reviews like these can
potentially dilute the impact of such negative sentiment terms in
reviews that do in fact report health hazards.

To investigate the impact of $s$, the number of negative reviews
sampled, we report in Figure~\ref{fig.nneg} the F1 score of the
informed prior classifier ($\tau=5$) as $s$ increases. We can see that
generally accuracy is stable for $s$ in the range $5,000-40,000$. For
values greater than 40,000, we suspect that the class imbalance
becomes too extreme for the instance weighting method discussed in
Section~\ref{sec.baseline}. For values less than 5,000, we may
actually have more positive than negative examples in the training
set, which is the reverse of what we expect in the unlabeled
data.

{\bf How often does the classifier identify potential product hazards
{\bf before} a recall is issued for a product?} Using the best
classifier from Table~\ref{tab.results} (informed prior; $\tau=5;
s=20,000$), we next predict the label for the reviews of the Amazon
products identified as being part of a CPSC recall (c.f.,
Section~\ref{sec.recall.db}). After filtering products with fewer than
10 reviews, we are left with 7,318 reviews from 86 products, of which
204 reviews were predicted to report a safety hazard.

To investigate the ability to provide consumers with a quicker
notification of potential hazards, Figure~\ref{fig.dayshist} shows a
histogram of when each of the identified hazardous reviews were
submitted to Amazon relative to the date that the product was recalled. We can
see that many of these reviews are posted well before the recall
date. There are some outliers appearing years before the recall date;
we observe that this can happen when a recall is issued because of
stores that continue to sell merchandise that had already been
recalled (in this case, not Amazon, but another
retailer). Additionally, there are reviews found well after the recall
date, which can occur for products that have been discontinued on
Amazon, but still have a page on which users can submit
reviews. Often, users post messages to warn others that the product
has been recalled.

As another view into this data, Figure~\ref{fig.leadtime} shows the
cumulative number of hazardous reviews found for each recalled product
over time (each line is a different product). This graph indicates
that, while for many products the classifier only identifies one or
two hazardous reviews, there are several products with five or more
hazardous reviews posted well before the recall date.

Additionally, Table~\ref{tab.recalls} lists five examples of hazardous
reviews identified before the recall date, ranging from 592 to 117
days prior. In many cases, the specific complaint described in the
Amazon review is also mentioned in the reason for the recall posted by
the CPSC. For example, the first row indicates a problem with the
front wheel assembly of a stroller, and the second row describes a
faulty adaptor in a car seat. The fourth review indicates that the
user has knowledge of a pending recall that had not yet been
announced.

Taken together, these results suggest that there is an opportunity to
mine Amazon reviews to provide earlier warnings to consumers about
potentially hazardous products.

{\bf How well does the classifier prioritize products to be
investigated for hazards?} Finally, we apply the classifier to all
915K Amazon reviews and compute indirect estimates of
accuracy. Overall, 10,857 were predicted to report a safety hazard. We
observe that the percentage of reviews classified as hazardous among
the non-recalled products is 1.2\%; for recalled products, it is
2.8\%. Thus, the classifier is more than twice as likely to classify a
review as hazardous for a recalled product, lending further support to
the accuracy of the model.

\begin{figure}[t]
  \centering
  \includegraphics[width=0.5\textwidth]{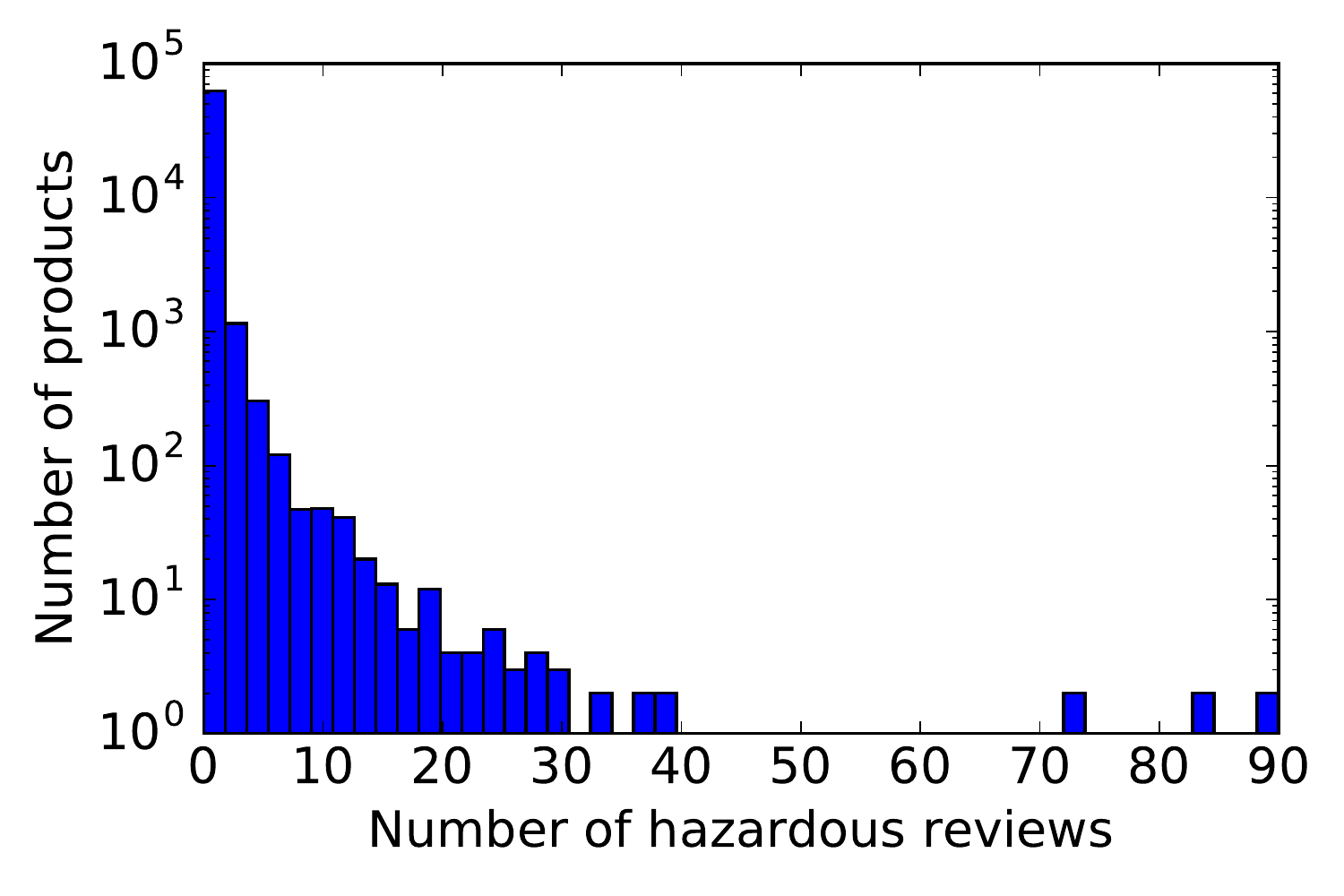}
  \caption{Number of reviews predicted to report a health hazard per product over all 915K Amazon reviews. \label{fig.pred_ratings}}
\end{figure}

Next, we count the number of hazardous reviews identified for each
product and plot the resulting histogram in
Figure~\ref{fig.pred_ratings}. We can see this follows the familiar
long tail distribution; there are a few products with many hazardous
reviews, and many products with a small number of hazardous
reviews. This matches our expectation that recalls are rare events.

As a final estimate of precision, we identify the ten reviews with the
highest posterior probability for the positive class. Of these, half
are reviews of recalled products; for the other half, while no recall
has been issued, the reviews contain strong language indicating that
further investigation may be required. For example, one review reports
of a child whose finger was stuck in a stroller, which is similar to
an incident that led to a recall of a different stroller. Similar
injuries are reported in the other reviews, and many contain phrases
liked ``This toy needs to be recalled ASAP!'' and ``Please do not buy
this product. It is unsafe!''

\begin{figure}[t]
  \centering
  \includegraphics[width=0.5\textwidth]{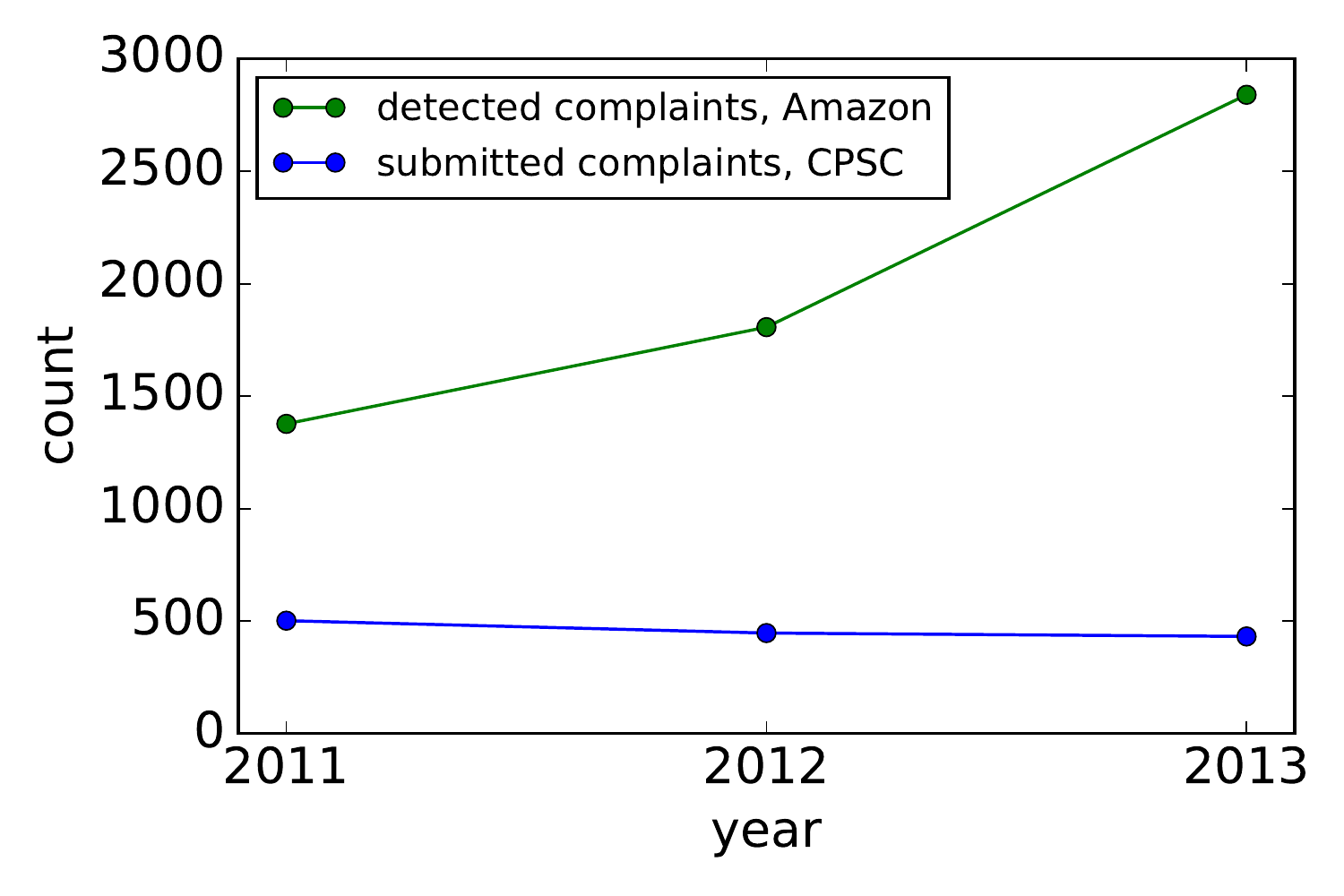}
  \caption{A comparison of the volume of consumer complaints in the ``Babies \& Kids'' category submitted to SaferProducts.gov versus Amazon reviews classified as reporting a health hazard in the same category. \label{fig.years}}
\end{figure}

To further determine how this approach may aid in the discovery of
safety hazards, we also compared the number of hazardous reviews
detected by year to the number of complaints submitted to the CPSC
over the same time frame. The number of hazardous reviews detected
indicates that many issues may be reported on Amazon, but not
submitted through the online portal at the CPSC. For example,
Figure~\ref{fig.years} shows that the classifier detected 2,840
reviews of baby products reporting health or safety hazards in 2013;
the CPSC complaint portal returns only 432 results for the same time
period. This suggests that consumers detect many more product hazards
than are submitted to CPSC (though it is possible some of these are
reported to the phone hot-line, which are not made public to our
knowledge). Furthermore, there is an inherent delay to the publication
of consumer complaints submitted to CPSC, due to the time required to
verify the complaint and contact the manufacturer. In our data, the
median time from when the report is submitted by a consumer to when it
is published on the CPSC website is 30 days. In contrast, Amazon
reviews are published immediately upon submission.

These results suggests that this system may be used to both discover
health hazards not submitted to the CPSC, as well as to prioritize
complaints posted on Amazon for potential examination for safety
hazards.

\begin{table*}[t]
\centering
\begin{threeparttable}
\begin{tabular}{|p{2.2cm}|p{1.7cm}|p{3.5cm}|p{1.7cm}|p{6.4cm}|}
  \hline
\textbf{Product Name} & \textbf{Recall Date} & \textbf{Recall Reason} & \textbf{Review Date} & \textbf{Review Snippet}\\
  \hline
Contours Options Tandem Stroller & 2012-07-24 & ``...the front wheel assembly can break, posing a fall hazard to the child in the stroller'' \tnote{a}  & 2010-12-10 (592 days prior) & ``...after less than 4 months of use, it fell apart: the front end collapsed because the two pins holding it in place popped out...I contacted Kolcraft immediately and nearly a month later I still don't have a working stroller.''\\
  \hline
Phil \& Teds Travel System Car Seat Adaptor & 2014-06-04 & ``the plastic adaptors used to connect an infant car seat to a stroller can crack, become unstable and break during use, posing a fall hazard to infants.'' \tnote{b} & 2013-04-30 (400 days prior) & ``I'm not sure if this attachment has a defect or if it is only supposed to have one button on the adapter, but it makes the carseat very wobbly and unstable...Is mine defective?  Everyone else seems to have great reviews, but mine is so unstable it seems dangerous.\\
\hline
Fisher-Price Rainforest Infant Swing & 2007-05-30 & ``infants can shift to one side of the swing and become caught between the frame and seat, posing an entrapment hazard.'' \tnote{c} &  2007-01-14 (136 days prior) &  ``It's a very poor design and needs a LOT of work.  And my daughter ends up in a crumbled up ball on one side of the swing more times than not.''\\
\hline
Phil \& Teds Dash Buggy Strollers & 2008-12-17 & ``the frame handle could fail to latch properly and break, posing a fall hazard to small children.'' \tnote{d} & 2008-08-18 (121 days prior) & ``The US distributor, Regal Lager is recalling all Dash strollers...The locking mechanism on the right side is defective and does not lock.''\\
  \hline
 Graco Activity Center & 2002-06-12 & ``The toy track can break, presenting a cut or pinch hazard and exposed small parts pose a choking hazard to young children.'' \tnote{e} & 2002-02-15 (117 days prior) & ``...the wheel part broke off of the activity part and left very sharp edges...he managed to get his head stuck in between the spoiler and the tray... (T)he parrimedics had to come...the walker had to be cut to release him.''\\
\hline
\end{tabular}
\begin{tablenotes}
\footnotesize
        \item[a] https://www.cpsc.gov/recalls/2012/kolcraft-recalls-contours-tandem-strollers-due-to-fall-and-choking-hazards
        \item[b] https://www.cpsc.gov/recalls/2014/philandteds-recalls-infant-car-seat-adaptors-for-strollers
        \item[c] https://www.cpsc.gov/recalls/2007/fisher-price-rainforest-infant-swings-recalled-due-to-entrapment-hazard
        \item[d] https://www.cpsc.gov/recalls/2009/regal-lager-recall-to-replace-phil--teds-strollers-due-to-fall-hazard
        \item[e] https://www.cpsc.gov/recalls/2002/cpsc-and-graco-announce-recall-of-toy-track-on-activity-centers
\normalsize
\end{tablenotes}
\caption{Examples of hazardous reviews identified prior to recall date. \label{tab.recalls}}
 \end{threeparttable}
\end{table*}

\section{Related Work}
\label{sec:related-work}

To the best of our knowledge, this is the first published system to
identify product health and safety hazards from online
reviews with no manual human annotation required. Additionally, our
time-series experiments indicate that these reviews can be identified
prior to the product recall date.

Very recently, \citenoun{winkler2016toy} used a keyword based approach
to identify online reviews that report injuries from toy
products. In addition to the manual effort required to curate
the keyword list, the approach appears to produce low precision rates
(9-44\%, depending on subcategory). Of the top 100 identified reviews,
only sixteen mentioned an injury.  The authors apply the same approach
to detect defects in dishwashers, with similar precision
values~\cite{law2017automated}. In contrast, our proposed approach
fits a statistical classifier with no human intervention required,
resulting in $>85\%$ precision and $>80\%$ recall.

Other recent work has identified vehicle defects in consumer reviews
using standard text classification, with accuracies ranging from
62\%-77\%~\cite{abrahams2015integrated}. However, in many domains it
is not feasible to annotate sufficient messages to use standard
supervised learning. Additionally, \citenoun{zhang2015predicting}
built an unsupervised approach to clustering vehicle defects by
subcategory. Such a method may serve to complement our present work by
providing more fine-grained clusters of reviews by hazard type.


\section{Conclusion}
We have presented a classification system to identify product reviews
on Amazon.com that indicate a health or safety hazard. The classifier
is trained without any additional human annotation or intervention by
using the consumer complaint records submitted to
SaferProducts.gov. To deal with data selection bias, we introduced a
new domain adaptation approach that is easy to implement and results
in an 8\% absolute increase over the best competing baseline. An
analysis of the historical reviews of recalled products indicates that
the system can identify potential safety hazards well before the
recall is issued.

In future work, we plan to build a web interface to make real-time
predictions as reviews are submitted to Amazon and to produce a ranked
list of potentially hazardous products. Additionally, we plan to
investigate classification methods that assign a severity to the
reported hazard, to further help consumer groups prioritize
investigations.

\bibliographystyle{aaai}
\bibliography{recalls}{}

\end{document}